# Multi-layer Ti-based Coating Obtained by Arc PVD Method.

K. Pavlov [#], K. Gorchakov [#], S. Gorchakova [#], K. Salojoki [#], V. Barchenko*, A. Sokolov ^

[#] - Laboratory of Kraftonweg Oy, Hiihtajantie 13 B, 49420 Hamina, Finland

* Saint Petersburg Electrotechnical University "LETI", ul. Professora Popova 5, 197376 St. Petersburg, Russian Federation

^ St. Petersburg State Polytechnical University, 195251, Saint-Petersburg, Polytechnicheskaya, 29, Russian Federation

*Abstract*

*We report the obtaining and primary studies of ~ 250µm thick multi-layer Ti-based protective coating deposited at high cooling rate from substance generated by cathode arc discharge in vacuum. High adhesion to steel substrate was attained through prior Arc plasma generator cleaning and successive Ion Bombardment method. All three arc-generated fractions including mainly droplet, vapour and ions have been utilised to form the coating. Obtained coating features pore-free, least defects and high hardness which, besides N presence, supposed to result from mainly martensitic transformations occurred at the presence of $N_2$. Two intermediate thin layers of Cu of few microns were achieved to insert within coating presumably to reduce overall elastic modulus of the material.*

**Introduction**

The well known Arc PVC method is widely used in research and industry for producing extra-hard and comparatively thin TiN-based films deposited from ion fracture of plasma [1][2][3]. These films are being thoroughly studied [4] [5] [6] and find their applications where extra hardness and wear resistance are crucial, particularly as cutting edges. While some application would need to combine high hardness and more thickness of up to 50-500 microns.

In presented experimental work the successive attempt to form hard Ti-Zr coating in thickness of 200-300 microns was made through deposition of entire mix of ions, vapor and micro droplets generated by arc-discharge, as well as the flexibility to form hard multi-layer coating in different metals is proved .

This article also presents the test results of physical and chemical properties of obtained Ti-Zr + Cu coating.

Corresponding authors:
[#] Konstantin Pavlov, Corresponding author. E-mail: *konstantinpavlov@hotmail.c*om
[##] Konstantin Gorchakov, Corresponding author. E-mail: *info.krafton@gmail.com*;
*Ph: +358 44 770 9016*

**Obtaining technique**

Both electrolytic-tough-pitch copper (Cu) and titanium-zirconium (Ti – Zr) alloy was chosen for manufacturing two cathodes as the materials to evaporate and be deposited. Blade-like stainless steel AISI 316 specimens were used to deposit the coating on. Specimens initially were cleaned, washed up in organic solvent and compressed air dried, then they were fixed in planetary rotating clamps inside vacuum chamber presented on Fig 1.

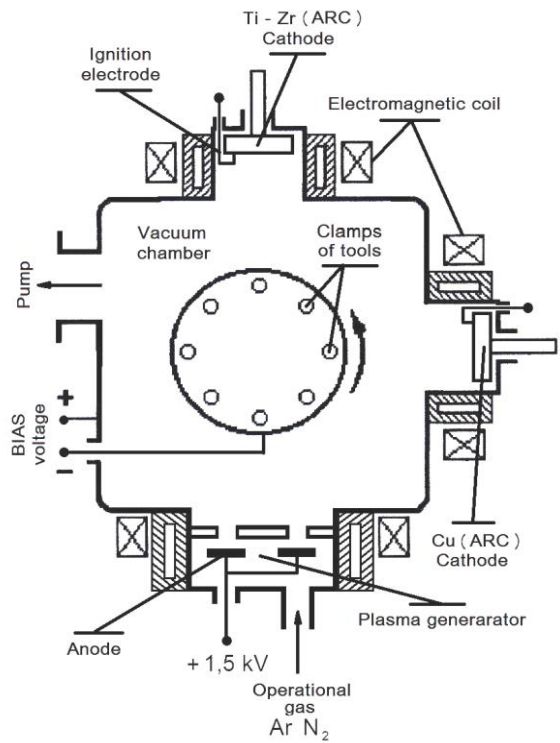

Fig. 1. Schematic drawing of chamber for Arc-method applied.

On achieving operational vacuum of $10^{-3}$ Pa the source of gas plasma was involved providing uniform gas discharged plasma in concentration of $10^{10} cm^{-3}$ and ion saturation current density of 10 mA/cm$^2$, besides, the negative bias voltage of 1500 V was applied to specimens. What made this initial stage to be responsible for substrate high cleaning through eliminating remaining oxides, films, contaminations, adsorbed gases and residual particles on substrate surface [7]  The next preparatory stage included ion bombardment process [8] to create the adhesion layer on the substrate  employing arc-generator of metallic Ti-Zr plasma operating at: arc current of 110 A;  operational  vacuum of $10^{-3}$ Pa; bias voltage applied to specimens of minus 1500 V.

Further stages were dedicated entirely to the formation of multi-layer composite  Ti-Zr-Cu coating. Primarily, the 1-st layer in thickness of about 100μm in Ti-Zr was deposited on adhesion surface at: specimens bias voltage  minus of 50V;  Ti-Zr cathode arc current of  110A; nitrogen N$_2$ was supplied increasing pressure in vacuum chamber of up to 2.5 x $10^{-2}$ Pa. Secondarily, Cu cathode plasma source was activated to form thin Cu layer in thickness of few microns at: specimens bias voltage of minus 50V; arc current of 75A; nitrogen was not supplied, pressure $10^{-3}$ Pa.  The next 3 layers were reproduced alternately at the same above modes. Total coating thickness of about 250μm has been reached.

The obtained composite coating have been examined using X-ray diffraction (XRD), optical metallography and scanning electron microscopy (SEM) as well as Vickers micro and normal hardness tests have been carried out.

The coating micro-structure was observed on its cross-sections with both MMP–4 optical microscope and TESCAN VEGA 5130 LM scanning electron microscope.

Vickers micro-hardness was tested on cross-section through PMT-3M tester with indenter loaded 20 g, while overall hardness was tested at 10 kg applied to indenter. The data was processed by Thixomet Pro image analysis system [9].

The phase structure was studied on DRON-1 diffractometer using X-Ray Structural Analysis technique.

**Results and Discussion**

The cross-section of coating was cut at about 16 degrees to substrate plane and photomicrograph (Fig. 2) was made with optical microscope.

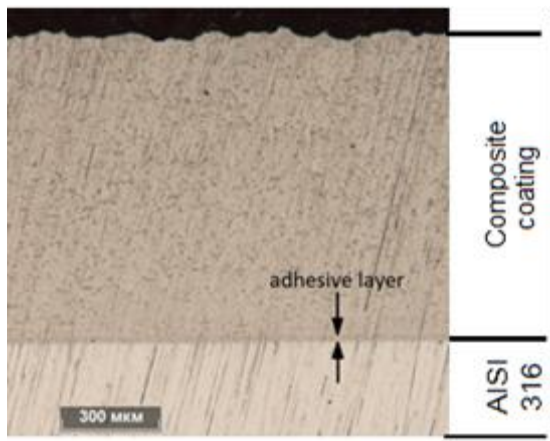

Fig. 2. Photomicrograph of coating cross-section cut at 16°.

SEM image of another cross-section cut at 90 degrees is shown on Fig. 3.

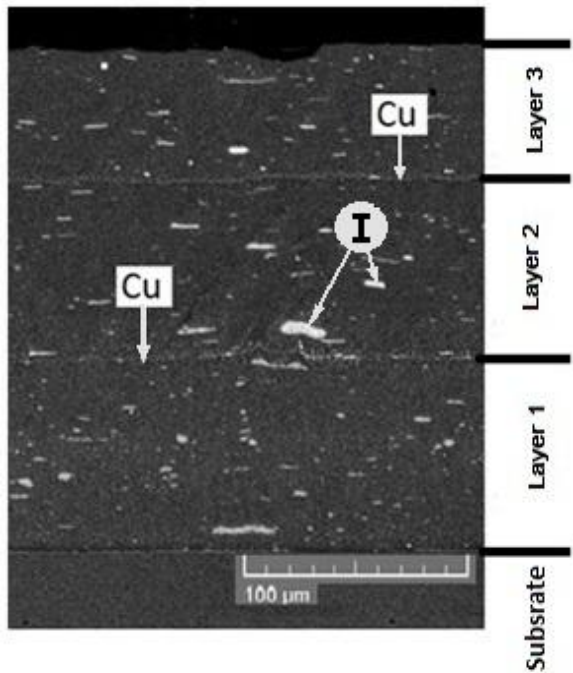

Fig. 3. SEM image of coating cross-section cut at 90°

The total composite coating layer of about 250 µm in thickness contains three main Ti-Zr layers and two thin Cu layers, all are distinctly presented. Careful visual examination of pictures revealed no cracks, cavities and lamination as well as no other similar defects within deposited Ti-Zr material. Adhesive layer and two thin intermediate Cu-layers were also found to be free from above defects. Depositing intermediary Cu layers was deemed to reduce elastic modulus (Young Modulus) of entire composite coating while the hardness of Ti - Zr layers to remain high. The micro-hardness tests (Fig.4) showed its values were decreasing with shifting to the substrate eventually descending to the normal one of substrate material (stainless steel AISI 316) $HV_{10}=165$ on average.

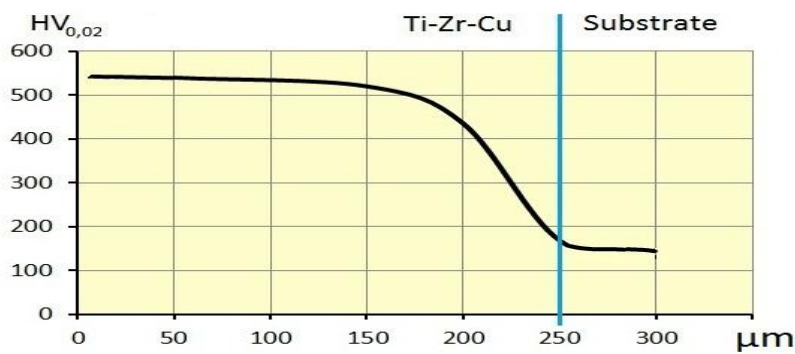

Fig. 4. Vickers micro-hardness tests versus depth of the coating.

The overall Vickers hardness was measured on coating surface and its value reached $HV_{10}=830$. Element quantity ratios in each layer of both obtained coating and substrate material are compared in Table I.

Tab. 1. Atomic percentages within the coating excluding Cu layers.

| Layer | C | Si | Ti | Cr | Mn | Fe | Cu | Zr | Total % |
|---|---|---|---|---|---|---|---|---|---|
| Layer 1 | | | 71,48 | | | | | 28,52 | 100.00 |
| Layer 2 | | | 71,31 | | | | | 28,69 | 100.00 |
| Layer 3 | | | 71,27 | | | | | 28,73 | 100.00 |
| White inclusion | | | 44,74 | | | | 1,23 | 54,03 | 100.00 |
| Substrate | 2,78 | 0,37 | | 13,82 | 0,61 | 82,42 | | | 100.00 |

Ⓘ — White inclusion

The atomic percentages of all three main layers have been discovered to be the same and fully corresponding to Ti - Zr ratio of evaporated cathode material. Discrepancy in atomic percentage between cathode and main coating materials has been determined only in oblong white inclusions (marked by Ⅰ) discernible within thick layers, what is supposed to happen owe to local inhomogeneity of evaporated cathode material, i. e. non-complete dissolution of Zr within Ti when cathode material had been alloyed. X-ray structural analysis of Ti–Zr layers revealed partially TiN and $Ti_2N$ compounds as well as $TiN_x$ substance with unknown stoichiometry, which could be determined only through crystal lattice parameters investigation.

Volume fraction of TiN in the main coating was detected to be not more than 3-5%.

Fig. 5 shows X-ray diffraction spectrum of main Ti–Zr layers of the coating a) and of cathode material used to form these layers b).

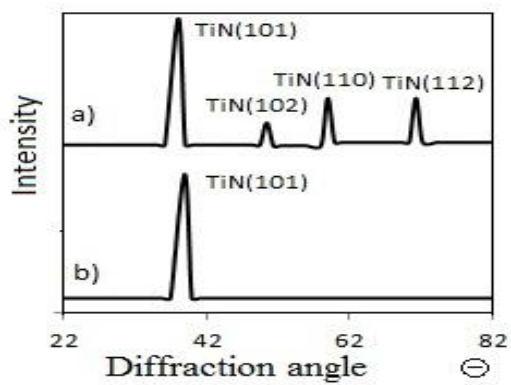

Fig. 5. X-ray diffraction spectrum of a) Ti-Zr layers and b) corresponding cathode material.

The cross-section of the coating was pickled by HF + $HNO_3$ mix acids solution in order to detect martensite structure. Series of strips spread across main Ti-Zr layers at oblique angle are particularly visible (Fig. 6) in the second layer while micro-hardness tests have revealed the highest values namely of the 3rd and 2nd layers.

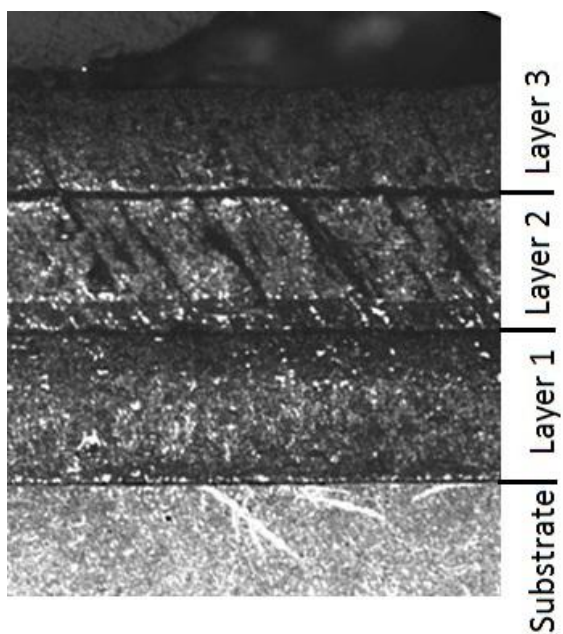

Fig. 6. SEM image of coating cross-section after pickling.

Thus, we believe that martensitic type of transformation in Ti-Zr structure took place in the presence of nitrogen [10, 11]. α - β phase transformation in nitrated layers (probably without other specific phase formations which have not been established) result in high hardness. Revealed structure evidences also rather high cooling rate at the surface what is confirmed by visible quenching zones in subsurface of substrate [12] [13].

Normally arc-plasma generates plasma stream constituting of three fractions: ion, vapor and molten droplets phases. Arc evaporated metal plasma flow was not subjected to any separation in our depositing process and therefore all fraction were used to form the coating under consideration and apparently every of them has contributed to general properties of the coating.

## Conclusions

1. Original technique of forming thick extra hard protective coatings (in thickness from 250μm and up) based on Arc evaporation is proposed.

2. Multi-layer composite coating in thickness of ~250μm with surface hardness attaining $HV_{10}=830$ and featuring practically no defects has been obtained on AISI 316 substrate.

3. The droplet fraction has been found to be mainly the coating was formed of.

4. Being rather simple and effective the proposed technique could be extensively used to improve operational properties of components surface made of steel and other metals.